\documentstyle[editedbook,epsfig,psfig,epsf]{mq}

\def\etal{{\em et al.} }

\def\cm2{cm$^2$ }
\def\se1{s$^{-1}$ }

 \begin{opening}
 \title{A Massive Jet-Ejection Event from SS433}
 \author{T. Kotani$^{0,1,2}$, S. Trushkin$^3$, E. K. Denissyuk$^4$} 
\institute{$^0$ NASA/Goddard Space Flight Center, Greenbelt, MD 20771, USA.\\
$^1$ U.\ of Maryland Baltimore County, 1000 Hilltop Circle, Baltimore,
 MD 21250, USA.\\
$^2$ Tokyo Inst.\ of Technology, Oookayama 2-12-1, Tokyo 152-8551, Japan.\\
$^3$ Special Astrophysical Observatory RAS, Nizhnij Arkhyz,
 Karachaevo-Cherkassia 369167, Russia.\\
$^4$ Astrophysical Institute of  Kazakh Academy  of  Sciences,
480068 Alma Ata, Kazakhstan.}
\author{N. Kawakita$^5$, K. Kinugasa$^5$, 
 S. Safi-Harb$^6$, \& D. Band$^{0,1}$}
\institute{$^5$ Gunma Astronomical Observatory, Nakayama 6860-86, 
Takayama, Agatsuma, Gunma 377-0702, Japan.\\
$^6$ U. of Manitoba, Winnipeg, MB R3T 2N2, Canada.}

\end{opening}
\runningtitle{A Massive Jet-Ejection from SS433}
\runningauthor{Kotani et al.}

\begin{document}
\vspace{-0.5cm}
\begin{abstract}
{\small The detection of a massive jet-ejection event from SS~433 with
 RXTE is reported.   SS~433 in its high state has been monitored with RXTE from
2001/11/09 (MJD = 52222) to 2001/11/25 (MJD = 52238), following a
radio flare on 2001/11/02 (MJD = 52215).  An irregular temporal
variation with time scales of $10^2-10^3$ s appears in the light
curve, and the amplitude increases day by day.  This is the first
detection of such a fast variation from the source.
In addition to the fast variations, the daily light curve 
scatters with a time scale of $\sim$day from 2001/11/17 (MJD =
52230).  Following the scatter, another radio flare has been detected
on 2001/11/22 (MJD = 52235), which has been obviously formed during the
X-ray scatter.  This is a preliminary report on a massive
jet-ejection event witnessed in X-ray band for the first time.
}
\end{abstract}

\section{Introduction}
The famous microquasar SS~433 shows two distinctive states; the
quiescent state in which the continuous jet flow is emanated, and the
high state in which massive jet blobs are successively
ejected\cite{fiedler87}.  While the former state has been well studied
with numerous X-ray missions, few massive jet-ejection events were
observed in X-ray band, except for a possible snapshot or two taken with
{\it Einstein}\cite{band89} and RXTE \cite{safi-harb02,band02}.
Because the ejection of a massive jet blob is a rare (2.6 yr$^{-1}$) and
short (a few days) event, it is difficult to observe with an X-ray
mission unless the observation is specially coordinated for that
purpose.

The situation is same for other microquasars such as GRS~1915+105.
While minor jet-ejection events in GRS~1915+105 have been observed with
RXTE in multi-wavelength campaigns and several other occasions (e.g.,
\cite{mirabel98,greiner96}), there are few X-ray observations of a
massive ejection event like the one reported in the famous paper by
Mirabel \& Rodir\'{\i}guez\cite{mirabel94}.  X-ray data of a massive
jet-ejection event in microquasars are needed to fill the void in our
understanding.  With such data, we can measure the mass of the massive
blob, determine the energy budget of the system, etc.

We planed TOO monitoring observations of SS~433 with RXTE\@.  In the
plan, a long-term monitoring is triggered with a radio flare, which
indicates that the source enters its high state.  In the high state, the
source is expected to experience a second flare within 30 days.  If the
second flare occurs, we can observe it from the onset.  Our plan is
different from the common TOO strategy for transient sources aiming at
the first flare.  The formation of a massive jet blob would not be
observed with an X-ray pointing observation after the detection of the
flare.  SS~433's radio activity is monitored with the RATAN-600 radio
telescope, which performs such a several-month-long monitoring
observation of SS~433 occasionally.

\section{The Observations and the Data Reduction}
\begin{figure}[h]
 \centering
\epsfig{file=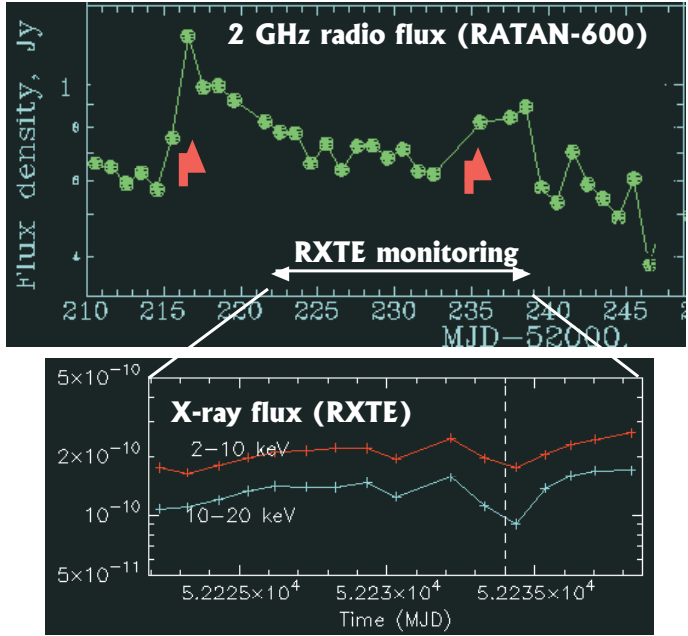,width=0.67\textwidth}
\caption{Radio and X-ray light curves of SS~433}\label{fig:xradio}
\end{figure}

\begin{figure}[h]
 \centering
\epsfig{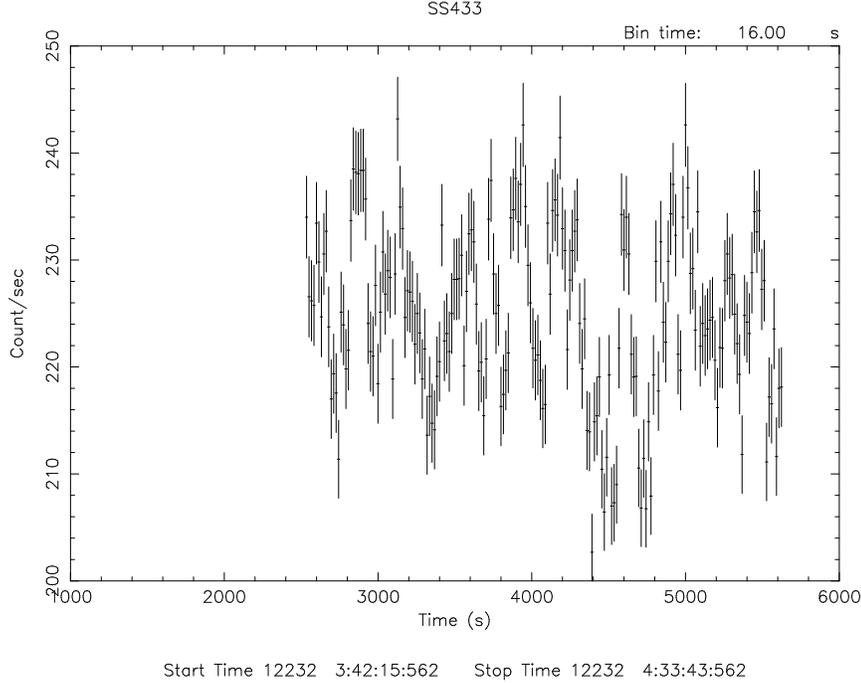}
\caption{X-ray light curve}\label{fig:blowup}
\end{figure}

A radio monitoring observation of SS~433 with RATAN-600 started on
2001/09/15.  After two months of static activity, a remarkable flare
occurred on 2001/11/02 (MJD = 52215).  The radio and X-ray light curves
are shown in figure~\ref{fig:xradio}.  Flux densities reached 1.3 Jy at
2.3 GHz at MJD = 52216.6\cite{kotani01}.  We triggered RXTE monitoring
observations as shown in Table~\ref{tbl:log}.  Except for a break at
2001/11/18, the source was observed for 3 ks everyday until 2001/11/25.
Both anodes of layer~1,~2, and~3 of the PCA have been used in the
Standard-2 mode.  The raw count rate of a data segment taken on
2001/11/19 is shown in figure~\ref{fig:blowup}.  Irregular variations
are clearly seen.  The spectrum and the response matrix have been made
from each segment and fitted with the traditional model used for the
data of SS~433 taken with Ginga and RXTE (e.g., \cite{kawai89,band02}):
\begin{equation}
\exp [ - \sigma N_{\rm H}] \times \{ \mbox{Bremsstrahlung} +
\mbox{thin Fe line} + \mbox{thick Fe line} \}.
\end{equation}
The hydrogen column density $N_{\rm H}$ is fixed to $0.6\times 10^{22}$.
From the fit, BGD-subtracted X-ray light curves are obtained as shown in
figure~\ref{fig:xradio}.  For more detailed spectroscopy,
see~\cite{safi-harb02}.

\begin{table}[h]
\centering
\begin{tabular}{ccc}
\hline
Start (MJD)	&End (MJD)	&PCU\\
\hline
2001/11/09 07:10 (52222.299)	&2001/11/09 08:11 (52222.341)  &0234\\
2001/11/10 05:19 (52223.222)	&2001/11/10 06:21 (52223.265)  &0234\\
2001/11/11 06:47 (52224.283)	&2001/11/11 07:47 (52224.325)  &0234 \\
2001/11/12 06:35 (52225.275)	&2001/11/12 07:35 (52225.316)  &0234 \\
2001/11/13 04:46 (52226.199)	&2001/11/13 05:46 (52226.240)  &0234 \\
2001/11/14 06:11 (52227.258)	&2001/11/14 07:13 (52227.301)  &0234 \\ 
2001/11/15 06:00 (52228.250)	&2001/11/15 07:01 (52228.293)  &0234 \\ 
2001/11/16 07:26 (52229.310)	&2001/11/16 08:32 (52229.356)  &023   \\
2001/11/17 07:14 (52230.302)	&2001/11/17 08:20 (52230.349)  &023   \\
2001/11/19 03:41 (52232.154)	&2001/11/19 04:33 (52232.190)  &0234  \\
2001/11/20 06:40 (52233.278)	&2001/11/20 07:46 (52233.324)  &02   \\ 
2001/11/21 08:05 (52234.337)	&2001/11/21 09:16 (52234.387)  &023  \\ 
2001/11/22 07:54 (52235.330)	&2001/11/22 09:05 (52235.379)  &024   \\
2001/11/23 04:32 (52236.189)	&2001/11/23 05:29 (52236.229)  &024   \\
2001/11/24 01:11 (52237.050)	&2001/11/24 01:54 (52237.080)  &02    \\
2001/11/25 05:45 (52238.240)	&2001/11/25 06:47 (52238.283)  &012\\
\hline
\end{tabular}
\caption{Observation log}
\label{tbl:log}
\end{table}

\section{Discussion}
The X-ray emitting part of the continuous jet flow with a speed of 0.26
c was estimated to be as long as $\sim 10^{13}$ cm, and the cooling time
of the X-ray emitting plasma as short as $\sim 10^3$ s~\cite{kotani97}.
According to this picture, an X-ray variation faster than $10^3$ s is
difficult to detect.  So, the discovery of the fast variation clearly
seen in figure~\ref{fig:blowup} is surprising, and our understanding of
the system must be somehow changed.  The fast variation suggests that
either the X-ray emitting part of the jet was as short as $\sim 10^{12}$
cm, or other source than the jet, e.g., the inner part of the accretion
disk or the surface of the compact object was seen.  To distinguish the
two possibilities, a temporal analysis of the Doppler-shifted iron
lines is quite promising.  If the jet itself was flickering, the iron
line would also show flickering, and if the variation was caused by some
instability of the accretion disk, the iron line would be stable through
the variation.  This report is preliminary and spectrally-resolved
temporal study is not yet  done.  The data taken on 2001/11/19 has been
divided into two, data with count rate larger than 220 counts s$^{-1}$
and with count rate less than 220 counts s$^{-1}$, and a spectrum has
been made from each data.  The fit results are not different
significantly.  So far, there is no evidence of an X-ray source other
than the jet.  And from the quick-look analysis, it can be said that the
fast variation is not periodic; no pulsation or QPO has been found from
the power spectra.  We can not determine whether the variation comes
from the accretion disk or a neutron star at this stage of analysis.

If we assume that the fast variation is resulted from the sudden change
of the power of the jet, the parameters of the jet would be derived.
The shortest time scale of the fast variation seems to be $\sim$50 s,
i.e., $4\times10^{11}$ cm, which corresponds to an initial electron
density of $10^{14}$ cm$^{-3}$, assuming an initial temperature of 20
keV~\cite{kotani97}.  Assuming a X-ray luminosity of $6\times10^{35}$
erg s$^{-1}$, the mass outflow rate of the jet would be $10^{-6}$
M$_\odot$ yr$^{-1}$, or a kinematic luminosity of $10^{38}$ erg
s$^{-1}$.

It should be noted that these estimated parameters are rather smaller
than those based on the data taken in the quiescent state.  Is the jet
weaker in the high state than in the quiescent state?  Or the previous
estimations are simply wrong and the jet is always short and weak?
Further analysis is going on.

\end{document}